# Community Bank Establishment and Consumption Growth: Evidence from Panel Study of Income Dynamics in USA

LI Yan

The Chinese University of Hong Kong

December 2024

**Abstract**

Consumption is a primary source of economic growth and key indicator of poverty. The establishment of community banks can provide credit resources, unlocking household consumption potential and playing a crucial role in economic development. This study explores the role of community banks in promoting consumption by using data from the Panel Study of Income Dynamics (PSID) for 11 waves from 1980 to 1990, and constructing a fixed-effects model using the time-varying difference-in-differences (DID) method. The findings indicate that the establishment of community banks effectively stimulates growth in local household consumption, primarily by increasing household income and reducing precautionary savings. Therefore, both government and financial institutions should further promote the development of regional financial institutions and credit tools to promote economic growth.

# 1 Introduction

Influenced by adverse conditions like war and deglobalization, the global economy has faced pressures such as demand contraction, supply shocks, and weakened expectations in recent years. In this context, unlocking consumption potential, improving financial services, and expanding demand have become crucial for promoting economic development. Additionally, consumption is considered an essential measure of poverty (Meyer and Sullivan, 2012). Promoting consumption growth is beneficial for alleviating poverty and narrowing income disparities.

The relationship between the establishment of financial institutions and economic growth has been a long-standing topic of interest and debate in economics. Community banks, which were once mainstream financial institutions in the United States, play a significant role in the U.S. financial system and have thus received considerable attention. Characterized by total assets of less than $1 billion and a focus on serving local businesses and residents, community banks can enhance household income through credit support, reduce the willingness to save, and consequently promote consumption.

Community banks play a vital role in the U.S. financial system, and there is considerable research on this topic. First, existing studies primarily focus on community banking operational issues, such as competition (Marquis and Lounsbury, 2007), diversification (Stiroh, 2004), and mergers (Hughes et al., 2019). Additionally, the impact of community banks on small and medium-sized enterprises (SMEs) is also a key area of focus, with past research covering topics such as debt financing (Jagtiani et al., 2016), bank-enterprise relationships (Berger et al., 2014), and loan growth (Avery and Samolyk, 2004). Finally, in developing countries like China and Nigeria, the role of community banks in poverty alleviation and economic growth has been widely discussed (Lu et al., 2021; Ayadi et al., 2008). However, research on the impact of community banks on American households and the economy is quite limited, and their role warrants further exploration.

The research sample from the 1980s in the United States holds particular significance. On one hand, as one of the wealthiest countries in the world, the U.S. has a highly developed financial sector. The contributions of small financial institutions in the U.S. can serve as a practical reference for developing countries and impoverished regions seeking to promote economic growth and escape poverty. On the other hand, while community banks are no longer mainstream financial institutions in contemporary America, the U.S. can draw upon past conclusions to reinforce its support for community banks, potentially revitalizing the economy during current challenging times. Lastly, although many countries have small financial institutions similar to community banks, data on these entities is relatively scarce, particularly in developing countries. Using the U.S. as a sample allows for collecting large-scale data, addressing the limitations of previous research that often relied on case studies and leading to more universally applicable conclusions.

To address the above issues and conditions, based on the theories of credit access, liquidity constraints, and precautionary savings, this paper analyzes the relationship between the establishment of community banks and household consumption growth using a sample of U.S. households participating in the Panel Study of Income Dynamics (PSID) from 1980 to 1990. The findings indicate that the establishment of community banks can alleviate household liquidity constraints, reduce precautionary savings, and release consumption potential. Furthermore, we conducted a series of robustness checks and validated the transmission mechanisms related to income and savings.

This paper contributes to bridging the research gap. First, while existing research has primarily focused on the operational issues of community banks or their impact on small and medium-sized enterprises, our study supports the positive effects of community bank establishment on households, thereby enriching the relevant literature. Second, among similar studies, this paper provides a detailed examination of the impact pathways of community banks both theoretically and empirically for the first time.

The paper is organized as follows: the first section is an introduction, followed by a literature review. The third section describes the sample and the research model. The fourth section presents and discusses the results. The fifth section highlights the conclusions and practical significance and recognizes the limitations.

## 2   Literature Review

From the late 1970s to the early 1980s, the strict restrictions on interstate banking competition in the United States created a favorable environment for the operations of local small banks. Furthermore, the Federal Reserve's adjustments to monetary policy, which started in 1979, further increased public demand for credit from small financial institutions. Community banks emerged in this macroeconomic environment. Although there is no strict definition of community banks, following previous research, this paper defines them as banks with assets of less than $1 billion, operating solely within their state, and providing a full range of financial services.

Community banks play a crucial role in the functioning of the U.S. financial system and the broader economy. Due to issues such as information asymmetry, small businesses and individuals often struggle to obtain long-term financing (Udell et al., 1993). Community banks, constrained by statutory lending limits, can focus on providing loans to small businesses and residents. In 1980, community banks allocated 20% to 30% of their loan portfolios to commercial loans. In summary, the primary clients of community banks are local SMEs and residents (De Young et al., 2004), providing essential banking services in small towns and rural communities across the nation.

By concentrating on local financial markets, community banks can enhance access to credit in their areas. Previous research has demonstrated that credit support is critical in creating job opportunities, expanding production, and increasing welfare (Luan and Bauer, 2016). Credit is also essential for alleviating poverty and increasing income (Guirkinger, 2008). Therefore, the establishment of community banks increases household income by providing credit, which in turn boosts consumption.

Then, standard macroeconomic models indicate that uncertainty plays a significant role in consumption and savings decisions, as evidenced by the convexity of the marginal utility of consumption. An increase in uncertainty generates positive additional savings, known as "precautionary savings" (Leland, 1968; Sandmo, 1970). While precautionary savings can act as a buffer for households facing unexpected economic pressures, they also crowd out household consumption and further restrict liquidity. The credit provided by the establishment of community banks can enhance financial security for households, allowing them to rely on credit to cope with sudden economic challenges, thereby reducing the need for precautionary savings and increasing consumption.

Liquidity constraints are considered one of the main reasons why the permanent income model fails to explain household consumption behavior (O. P. Attanasio and Weber, 2010). In imperfect credit markets, households are unable to convert expected income into current consumption through credit Zeldes, 1989). Families anticipating an increase in income are forced to delay subsequent consumption growth Deaton, 1991). Therefore, the establishment of community banks can provide credit support, alleviating liquidity constraints for local households, which enables smoother consumption patterns and raises current consumption levels.

In conclusion, the theoretical framework supports the notion that increasing income and reducing saving can lead to increased consumption. This paper argues that the establishment of community banks can promote consumption growth among local households.

# 3 Methodology

## 3.1 Sample Selection and Data Source

To examine the relationship between the establishment of community banks and consumption growth, this study selects families who participated in the Panel Study of Income Dynamics (PSID) survey from 1980 to 1990 as the sample. The relevant data on family consumption and characteristics all come from the PSID. Staring from 1968, PSID is the longest-running longitudinal household survey in the world, which collects data from over 5,000 families on topics including employment, income, expenditures, health, and many other areas, providing an excellent national representation. The establishment year of community banks in each state is sourced from the Federal Deposit Insurance Corporation (FDIC), which is an independent agency created by Congress to maintain the stability of the nation's financial system. To ensure the validity of the research data, the following methods were employed for data screening and processing: 1) Hawaii and U.S. overseas territories were excluded; 2) Samples with household heads aged over the U.S. retirement age (66 years old) were excluded; 3) Samples with missing values and outliers were excluded. 4) Only families where the head of the family remained unchanged were retained to facilitate tracking. 5) Samples with zero consumption were excluded. The final sample consists of 41,374 observations from 5,479 families.

## 3.2 Variable Setting

Consumption (Com): The logarithm of Consumption is the dependent variable in this study. Prior to 1999, the PSID had limited surveys related to consumption. This paper draws on the previous research (Hall and Mishkin, 1982; O. Attanasio and Pistaferri, 2014), using food and rental expenditures as a proxy for overall household consumption. In the PSID, food expenditures are composed of expenditures at home, away from home, and food stamps. Given that food stamps are provided by the government and are less influenced by consumption preferences, this study only calculates the sum of food expenditures at home and away from home. Then, considering the PSID did not survey food expenditures in 1988 and 1989, we used the interpolation method to fill in the data for these two years. Additionally, we took 6% of the house's value as rental expenditure and added that to the food expenditure. Finally, We deflated the sample's consumption data based on the U.S. 1980 CPI.

Community Bank (Policy): The establishment of community banks is the independent variable in this study. While the definition of community banks is somewhat ambiguous, it generally refers to banks with assets under $1 billion. Furthermore, to ensure a more accurate estimation of the impact of community bank establishment on local consumption growth, we selected the establishment dates of the first community bank in each state from the FDIC. We ensured that these community banks operated solely within their respective states.

By drawing on previous research on consumption and the PSID (Gruber, 1998; Meyer and Mok, 2019), we selected several control variables to enhance the study's validity. The control variables mainly come from two levels: 1) Head level: the gender of the household head; 2) Household level: household size, household income, and average education level of head and wife.

The explanation and set of variables are in Table 1.

Table 1: Variables Description

| Symbol | Variable | Variable Definition |
|---|---|---|
| **Com** | Total consumption | Annual food and rental expenditure |
| **Policy** | Community Bank Establishment | Set to 1 if the state has already established the first community bank; otherwise 0. |
| **Edu** | Education level | Average education level of the head and wife |
| **Sex** | Gender of head | Which is set to 1 if head is male, otherwise 0 |
| **Income** | Total income | Actual family total income |
| **Size** | Family size | Total members of the family |
| **IND** | Individual-fixed effect | Family dummies |
| **STATE** | State-fixed effect | State dummies |
| **YEAR** | Year-fixed effect | Year dummies |

## 3.3 Methodological Remarks

To examine the relationship between the establishment of community banks and consumption growth, this study employs a time-varying difference-in-differences model with individual, state and year-fixed effects to construct Equation (1):

$$\text{Log(Consumption}_{j,c,t}) = \beta_0 + \beta_1 \text{Policy}_{c,t} + \beta_2 \text{Controls}_{j,c,t} + \omega_j + \epsilon_c + \phi_t + \mu_{j,c,t} \quad (1)$$

In this equation, Consumption$_{j,c,t}$ is the food and rental consumption of the family $j$ from state $c$ in year $t$, and we use its logarithm value. The variable Policy$_{c,t}$ is set to 1 if in year $t$ the first community bank has been established in state $c$. The term Controls$_{j,c,t}$ represents the control variables at two different levels, including educational level, gender, family income, and family size. $\omega_j$, $\epsilon_c$ and $\phi_t$ are individual(family)-fixed effects, state-fixed effects and year-fixed effects. $\mu_{j,c,t}$ is residuals. $\beta_1$ represents the effect of the establishment of a community bank on consumption. All data were clustered at the family-level. Detailed definitions of these variables are given in Table 1.

# 4 Results and Discussions

## 4.1 Descriptive Statistics

Descriptive statistics are shown in Table 2, which summarizes the distribution and characteristics of each variable analyzed in the study. The results show that the Mean of consumption and Log(consumption) are 4809 and 8.124, and the Mean of Policy is 0.438. The mean close to 0.5 indicates no significant difference in the scale of observations before and after the establishment of the bank. The maximum and minimum values of consumption are 0.827 and 51,490, respectively, indicating a significant difference in sample distribution and a high degree of differentiation. The same applies to the data on income.

Table 2: Descriptive Statistics

| Variable Name | Obs | Mean | SD | Min | Max |
| --- | --- | --- | --- | --- | --- |
| **Com** | 41,374 | 4,809 | 3,895 | 0.827 | 51,490 |
| **Log(Com)** | 41,374 | 8.124 | 0.917 | -1.189 | 10.849 |
| **Policy** | 41,374 | 0.438 | 0.495 | 0 | 1 |
| **Income** | 41,374 | 27,002 | 31,149 | 0 | 1,412,200 |
| **Edu** | 41,374 | 3.622 | 1.890 | 0 | 8 |
| **Sex** | 41,374 | 0.760 | 0.427 | 0 | 1 |
| **Size** | 41,374 | 2.946 | 1.541 | 1 | 14 |

## 4.2 Regression Result Analysis

Based on Equation 1, we systematically and progressively incorporated control variables from two distinct levels of analysis, with the comprehensive results presented in columns (1) to (3) of Table 3. This methodological approach allows us to better understand and illustrate the nuanced impact of the establishment of community banks on local consumption patterns.

In columns (1) through (3), the coefficients for the variable "Policy" consistently demonstrate positive values and are statistically significant at the 1% level. This strong statistical significance indicates that the establishment of community banks has a substantial and positive effect on promoting local consumption growth. The empirical findings we observed not only reinforce our theoretical inferences but also highlight the critical role that community banks play in enhancing economic activity within the region.

Focusing specifically on the results presented in column (3), we find that the annual food expenditure of households in the state is projected to increase by an average of 3.9% following the establishment of community banks. This increase in food expenditure reflects a broader trend of enhanced consumption growth.

Table 3: Regression Results

| VARIABLES | (1) | (2) | (3) |
|---|---|---|---|
| **Policy** | 0.047*** | 0.043*** | 0.039*** |
|  | (3.57) | (3.36) | (3.26) |
| **Sex** |  | 0.348*** | 0.311*** |
|  |  | (10.06) | (10.10) |
| **Edu** |  | 0.128*** | 0.051*** |
|  |  | (20.21) | (7.93) |
| **Income** |  |  | 0.000*** |
|  |  |  | (5.04) |
| **Size** |  |  | 0.154*** |
|  |  |  | (24.56) |
| **Indivi.-Fixed effect** | Yes | Yes | Yes |
| **State-Fixed effect** | Yes | Yes | Yes |
| **Year-Fixed effect** | Yes | Yes | Yes |
| **Constant** | 8.122*** | 7.397*** | 7.168*** |
|  | (1,433.10) | (283.65) | (279.41) |
| **Observations** | 41,374 | 41,374 | 41,374 |
| **R-squared** | 0.736 | 0.757 | 0.776 |

*Notes: t-statistics in parentheses \*\*\* p is less than 0.01, \*\* p is less than 0.05, \* p is less than 0.1*

## 4.3 Robustness Test: Parallel Test

To investigate the dynamic changes in consumption before and after the establishment of community banks and to ensure that the promoting effect of community bank establishment on consumption growth does not stem from pre-existing trends, this study draws on the methodologies of previous research (Jacobson et al., 1993; Kudamatsu, 2012). We employ an event study approach to conduct a parallel trends test, thereby enhancing the robustness of the results. Specifically, we set up the regression equation as follows:

$$\text{Log(Consumption}_{j,c,t}) = \alpha + \sum_{\substack{k=-5 \\ k \neq 0}}^{\geq} \delta_k D_{c,t}^k + \gamma \text{Controls}_{j,c,t} + \omega_j + \epsilon_c + \phi_t + \mu_{j,c,t} \quad (2)$$

In Equation (2), let the year that state $c$ establishes the first community bank be $s$, and $D$ is a dummy variable indicating the establishment of the community bank. When $t - s = k$, $D_{c,t}^k = 1$; otherwise, $D_{c,t}^k = 0$. To maintain a reasonable period, we define $D_{c,t}^3 = 1$ when $t - s > 3$ and $D_{c,t}^{-5} = 1$ when $t - s < -5$. The control variables, fixed effects, and residuals are the same as in Equation (1).

To avoid multicollinearity, this study uses the year of the establishment as the baseline year,

excluding the dummy variable for k = 0. Additionally, following the previous approach (Beck et al., 2010), this study performed de-trending and centering in the parallel trends test. The final results of the parallel trends test are illustrated in the Table 4 and Figure 1 below:

Table 4: Parallel Test Regression Results

| VARIABLES | (1) |
| --- | --- |
| Pre_5 | 0.009 |
|  | (0.60) |
| Pre_4 | -0.003 |
|  | (-0.84) |
| Pre_3 | -0.006 |
|  | (-1.25) |
| Pre_2 | -0.016 |
|  | (-2.19) |
| Pre_1 | 0.026 |
|  | (0.39) |
| Post_1 | 0.058** |
|  | (2.43) |
| Post_2 | 0.078*** |
|  | (2.89) |
| Post_3 | 0.099*** |
|  | (2.97) |
| Baseline Year | Omitted |
| Control Variables | Yes |
| Indivi.-Fixed effect | Yes |
| State-Fixed effect | Yes |
| Year-Fixed effect | Yes |
| Constant | 7.172*** |
|  | (264.20) |
| Observations | 41,374 |
| R-squared | 0.776 |

*Notes: t-statistics in parentheses \*\*\* p is less than 0.01, \*\* p is less than 0.05, \* p is less than 0.1*

As Table 4 and Figure 1 shows, in the pre-baseline period, the 95% confidence intervals all include 0, indicating that the explanatory variables' regression coefficients and trends are insignificant. However, in the post-baseline period, the regression coefficients of the explanatory variables are significantly positive and maintain an upward trend. This indicates that after the establishment of community banks, local household consumption significantly increases, confirming that the consumption growth is indeed driven by the establishment of the banks rather than a time trend effect.

Figure 1: Parallel Test

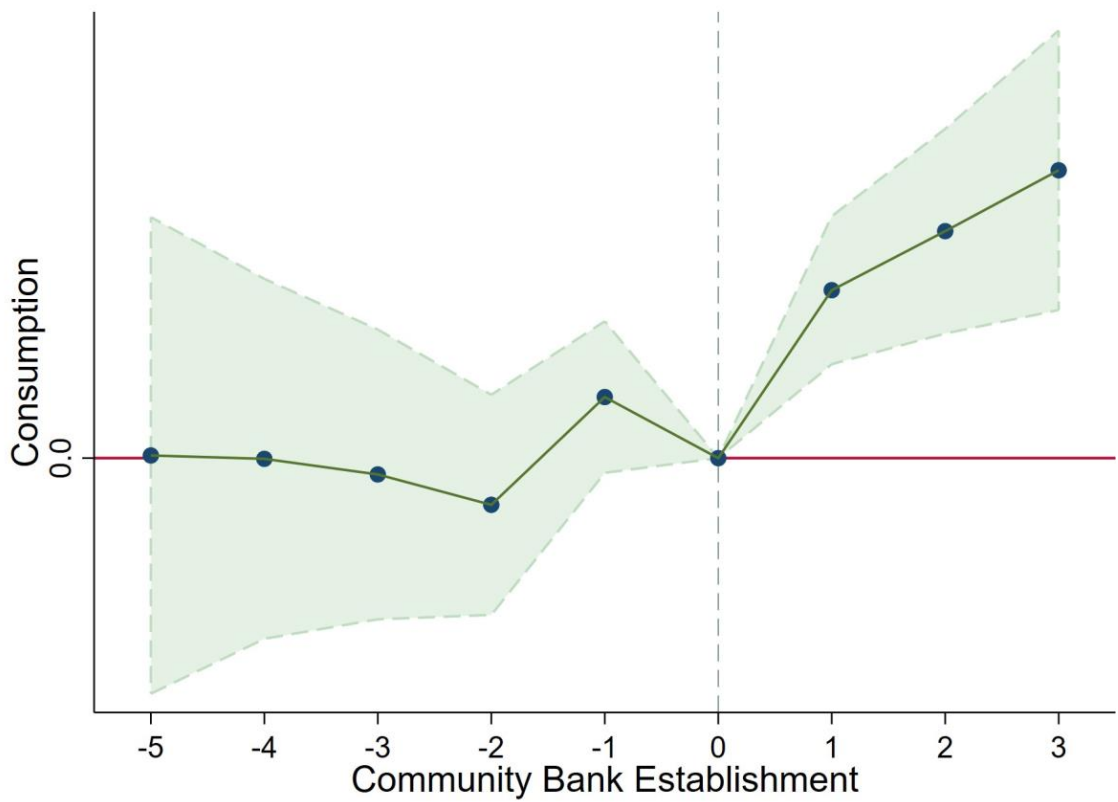

## 4.4 Robustness Test: Placebo Test

To rule out the influence of other concurrent events or unobserved characteristics of states, we conducted a placebo test. Specifically, we randomly designed the year of establishment of community banks in the sample and the corresponding region, thereby generating a dummy variable for the policy as a new explanatory variable in the regression analysis, and then performed the regression. This process was repeated 500 times to obtain the regression coefficients and density distribution of the placebo test. If other concurrent events drove the growth of household consumption, we would expect to see similar results after randomly assigning the establishment and location of community banks.

However, the findings presented in Figure 2 reveal that both the P-value and the density distribution conform to a normal distribution symmetric at x=0. Notably, the actual coefficient from the baseline regression, which is 0.039, exceeds all coefficients obtained from the placebo tests. In conclusion, the outcomes of the placebo test strongly suggest that consumption growth is not influenced by contemporaneous events or regional characteristics.

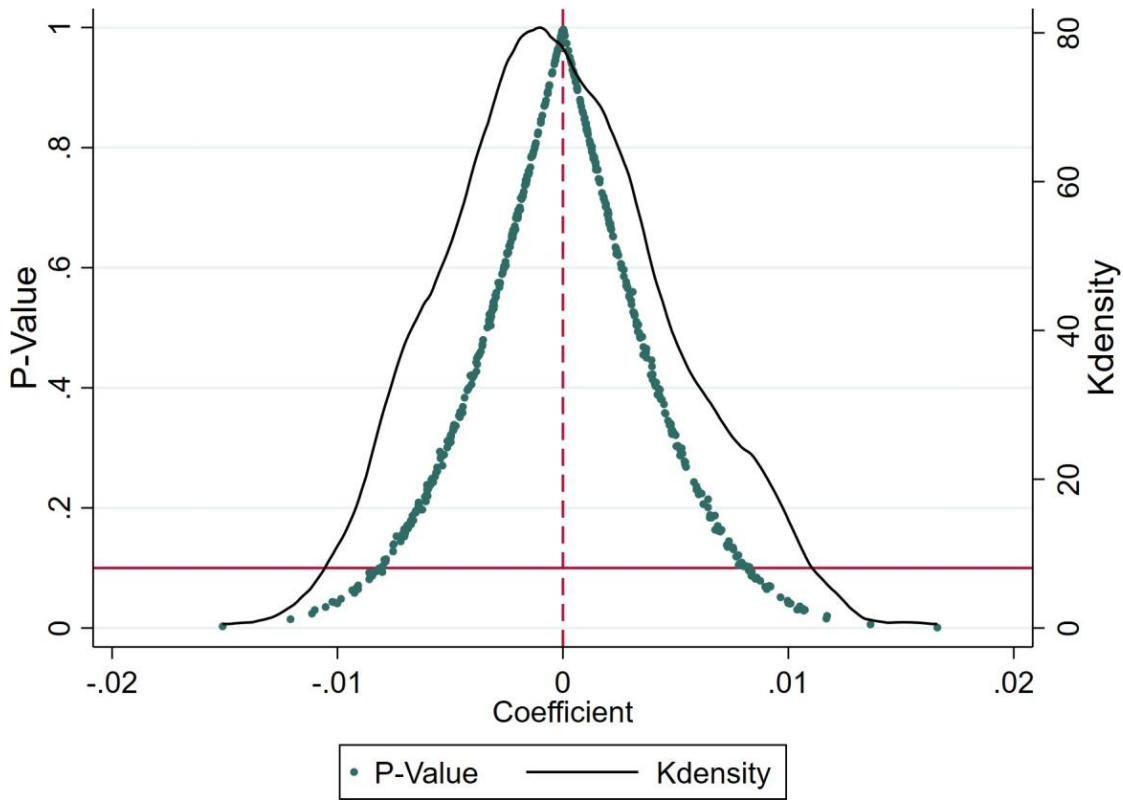

Figure 2: Placebo Test

## 4.5 Robustness Test: Adjustment of sample and variables

In order to make the estimation more accurate and reliable, this paper specifically adjusts the samples and variables and re-runs the regression.

First, considering that low-income households have a relatively strong demand for credit and a higher marginal propensity to consume food, the establishment of community banks is likely to have a more pronounced effect on the consumption of impoverished populations. Therefore, we restricted the sample to households with annual incomes above the U.S. poverty line for the regression analysis. The results are shown in column (1).

Second, due to the lack of food consumption data for 1988 and 1989 in the PSID, this study's baseline regression supplemented these two years using the interpolation method. To avoid any potential impact of the supplemented data on the accuracy of the estimates, we conducted the regression using only the original data (excluding 1988-1989). We replaced the consumption variable with only food expenditures, ensuring that the dependent variable underwent sufficient adjustment. The results are shown in column (2).

Third, due to the increase in interest rates in the late 1970s, the U.S. economy experienced a prolonged negative growth period from early 1980 to the end of 1981. Therefore, we excluded the data from 1980-1981 to ensure a relatively stable macroeconomic environment, and the results are shown in column (3).

Finally, the baseline regression includes rental expenditure and food expenditure into consumption. Considering that rental expenditure is highly sensitive to credit access, we only include food expenditure into consumption calculations. The results are presented in column (4).

The four robustness checks' results, conducted through variables and sample adjustments, are presented in Table 5. These findings indicate that the coefficient for Policy consistently remains positive and is significant at the 1% or 5% level, thereby affirming that the promotion of consumption attributable to the establishment of community banks is robust across various specifications.

Table 5: Robustness Test Regression

| Variables | (1) | (2) | (3) | (4) |
|---|---|---|---|---|
| Policy | 0.026** | 0.027*** | 0.034*** | 0.029** |
| | (2.39) | (2.15) | (2.87) | (2.48) |
| Control Variables | Yes | Yes | Yes | Yes |
| Indivi.-Fixed effect | Yes | Yes | Yes | Yes |
| State-Fixed effect | Yes | Yes | Yes | Yes |
| Year-Fixed effect | Yes | Yes | Yes | Yes |
| Constant | 7.313*** | 7.167*** | 7.225*** | 7.262*** |
| | (246.42) | (324.36) | (250.81) | (338.82) |
| Observations | 32,234 | 35,028 | 33,120 | 41,279 |
| R-squared | 0.787 | 0.678 | 0.798 | 0.690 |

*Notes: t-statistics in parentheses \*\*\* p is less than 0.01, \*\* p is less than 0.05, \* p is less than 0.1*

## 5 Analysis of Mechanisms

This study argues that the establishment of community banks promotes household consumption through two pathways: increasing household income and reducing household precautionary savings. Given the clear impact and strong causal relationship of income and savings on consumption, we draw on previous econometrics research (Jiang, 2022) to examine the effects of community bank establishment on household income and savings separately. This approach aims to verify whether these two factors serve as channels of influence.

First, we continue to utilize the PSID survey data, with household income as the dependent variable; the results are shown in column (1). Secondly, since the PSID only surveyed savings decisions in 1984 and 1989 from 1980 to 1990, we used data from these two years, with the results in column (2). In column (2), the savings decision is a binary variable: it equals one if the household responds affirmatively to both general savings (cash, bank accounts) and other savings (checking accounts, insurance, pensions, etc.); otherwise, it equals 0. The results in Table 6 indicate that the establishment of community banks increases total household income while also reducing the likelihood of households making savings decisions, and the results are significant at 5% level.

Table 6: Regression of Impact Mechanisms

| Variables | (1) Income | (2) Saving Decision |
| --- | --- | --- |
| **Policy** | 944.169** | -0.236** |
|  | (2.43) | (-2.31) |
| **Control Variables** | Yes | Yes |
| **Indivi.-Fixed effect** | Yes | Yes |
| **State-Fixed effect** | Yes | Yes |
| **Year-Fixed effect** | Yes | Yes |
| **Constant** | 81.67 | 0.171*** |
|  | (0.60) | (2.77) |
| **Observations** | 43,354 | 5,964 |
| **R-squared** | 0.677 | 0.640 |

*Notes: t-statistics in parentheses \*\*\* p is less than 0.01, \*\* p is less than 0.05, \* p is less than 0.1*

## 6 Discussion and Conclusion

The scale and number of community banks in the United States have significantly contracted under the influence of various factors such as policy restrictions, intensified competition, and financial digitalization. However, research on community banks remains highly relevant. This paper utilizes PSID surveys to obtain the establishment dates of community banks across various states. It draws on representative data regarding household consumption and income in the U.S. from 1980 to 1990 via the PSID to study the impact of community bank establishment on consumption growth. The findings indicate that the establishment of community banks can significantly promote consumption growth among local households. Subsequently, the paper strengthens the robustness of its conclusions through parallel tests, placebo tests, and sample modifications.

Finally, this study explores the mechanisms through which community banks affect consumption. It finds that, on the one hand, the establishment of community banks provides credit support, promoting economic growth by creating jobs and expanding production, thereby increasing residents' income and consumption. On the other hand, household uncertainty may encourage families to increase their expected savings; however, the access to credit provided by community banks can enhance households' ability to cope with sudden economic pressures, reducing the need for savings and releasing consumption potential. Furthermore, liquidity constraints make it difficult for households to convert expected income into current consumption, which contradicts the permanent income model. At the same time, the establishment of community banks can improve household liquidity, thereby smoothing and increasing consumption.

This paper has policy formulation, financial institution governance, and poverty alleviation implications. First, given the significant reduction in the number of community banks in the U.S., it may be worthwhile to consider policies that promote the development of community banks to alleviate economic pressures in the face of challenges. Second, impoverished regions or developing countries can draw lessons from the U.S. experience by using community banks to provide credit support for poverty alleviation efforts. Lastly, community banks can further enhance their support for local SMEs and low-income families, thereby maximizing their social value. Community banks might also consider expanding their customer outreach through digitalization and other reform measures.

The implicit problem of this paper is that the period of community banks establishment coincides with the U.S. savings and loan crisis. In the early 1980s, the special macroeconomic environment, policies and regulations gave rise to the rapid development of small financial institutions. The growth of community banks and S&Ls during this period was largely due to their issuance of bad loans and disorderly expansion without reasonable supervision. By the end of the 1980s, policy shifts and early risk accumulation led to a large number of bankruptcies of such institutions (Curry and Shibut, 2000; Kane, 1989). In our study, we studied the impact of the first community bank established in each state on consumption, but did not take into account that its impact on consumption may come from the special background of the times. In other words, its impact on consumption may not be due to the corresponding theoretical and mechanism analysis, and it may not be persistent with the bankruptcy of financial institutions. This problem affects the accuracy of the conclusion and weakens the reference significance of the research to other regions and the present.

Our study still has some other limitations. First, the lack of a unified definition for community banks complicates the identification of the first community bank in various regions, and inaccuracies in the establishment dates of community banks may affect the conclusions. Second, our research primarily focuses on data from the U.S. between 1980 and 1990, and given the substantial decrease in the number of community banks today, the practical significance of this paper is somewhat limited. Finally, while the paper finds that community banks hold additional significance for impoverished and developing countries, it does not conduct an in-depth study on this aspect.

Future research could explore the following areas: First, topics could be closer to current events, such as studying the negative impacts of community bank closures and mergers on consumption. Second, a deeper investigation could be into their effects on poverty alleviation and other household economic indicators. Third, the research scope could be further expanded to developing countries to seek more general conclusions. Fourth, More detailed consumption data than just food and rent expenditures can be used for further research, such as Consumer Expenditure Surveys. Lastly, the study should try to change the sample year to eliminate the impact of the U.S. savings and loan crisis at the same time on the robustness of the results.